\documentclass[preprint,3p,twocolumn,times,10pt,sort&compress]{elsarticle} 

\usepackage[running]{lineno}

\usepackage{float}
\usepackage{subfig}

\usepackage{braket} 
\usepackage{amsfonts}

\usepackage[numbers]{natbib}
\usepackage{amssymb}
\usepackage{amsmath}
\usepackage{graphicx}

\graphicspath{{Figures/}}

\usepackage{mathrsfs}
\usepackage{mathtools}
\usepackage{upgreek}
\usepackage{float}
\usepackage[dvipsnames]{xcolor}
\usepackage{framed}
\usepackage{mathtools,cancel}
\usepackage{siunitx}
\usepackage{textcomp}
\usepackage{xcolor}
\usepackage{listing}
\usepackage{bibentry}
\usepackage{hyperref}

\usepackage{AAS_TeX_Macros} 

\begin{document}

\begin{frontmatter}

\title{Removal of $^{210}$Pb by Etch of Crystalline Detector Sidewalls}

\author[SDSMT]{Joseph Street\corref{primary}}
\cortext[primary]{Corresponding author}
\ead{joseph.street@mines.sdsmt.edu}
\author[TAMU]{Rupak Mahapatra}
\author[SDSMT]{Eric Morrison}
\author[TAMU]{Mark Platt}
\author[SDSMT]{Richard Schnee}
\address[SDSMT]{501 E. Saint Joseph St., Rapid City, SD 57701}

\address[TAMU]{400 Bizzell St, College Station, TX 77843}



%
%

\begin{abstract}
A potential source of dominant backgrounds for many rare-event searches or screening detectors is from radon daughters, specifically $^{210}$Pb, deposited on detector surfaces, often during detector fabrication. Performing a late-stage etch is challenging because it may damage the detector. This paper describes a late-stage etching technique that reduces surface $^{210}$Pb and $^{210}$Po by $>99\times$ at 90\% C.L.
\end{abstract}

\begin{keyword}
\texttt{dark matter, low-background searches, crystal, etching, radon, contamination, surface, Pb-210}
\end{keyword}

\end{frontmatter}


\section{Introduction}
Radioactive contamination on detector surfaces can limit the sensitivity of rare-event searches or low-background measurements. A potentially dominant source of surface contamination is from radon progeny, such as $^{210}$Pb, which has a 22.3 year half-life, long compared to the lifetime of most experiments. These progeny produce several decays that can mimic a signal in these detectors~\cite{Simgen:2013dlh,Bunker:2014bea,Schumann:2019aa,snowmass2013screening,cuore2011backgrounds,LRT2004Leung,xmassLRT2010}.

SuperCDMS is a rare-event, dark-matter search experiment that uses high-purity Si or Ge crystals as the target material~\cite{Agnese:2016cpb}. Sensors on the faces of these cylindrical detectors measure athermal phonons (heat) and ionization (charge) produced from particle scattering off Si or Ge nuclei. Primarily focused on detecting the elusive weakly interactive massive particle (WIMP), SuperCDMS has produced world-leading exclusion limits on the WIMP-nucleon scattering cross-section~\cite{Agnese:2017jvy,Agnese:2016aa}, and its detector technology has evolved to be sensitive to even single electron-hole pairs~\cite{Agnese_2018,2018ApPhL.112d3501R}.

This paper describes a newly-developed etching technique for the sidewalls of solid-state crystal detectors. A Si crystal core was exposed to high-radon air, depositing $^{210}$Pb on the core's surface. The core's sidewall was assayed by an alpha counter sensitive to $^{210}$Po alpha decays before and after the etching technique was performed to infer the reduction in $^{210}$Pb due to the etch.

\section{$^{210}$Pb Deposition from High-Radon Air Exposure}
A 76\,mm $\times$ 25.4\,mm polished Si crystal was fabricated with SuperCDMS sensors. Following fabrication, the crystal was coated with Logitech OCON-200 bonding wax and sandwiched between two borosilicate glass disks to protect the circuit and reduce substrate chipping during coring. The coring process was performed on a standard Bridgeport Mill using hardware custom-designed to aid in substrate cooling. The diamond-coated coring bit ($1.000\pm0.005\,$in diameter) was custom-fabricated by Keen-Kut Products, Hatward California.

\begin{figure}[t]
	\centering
  	\includegraphics[width=1\linewidth]{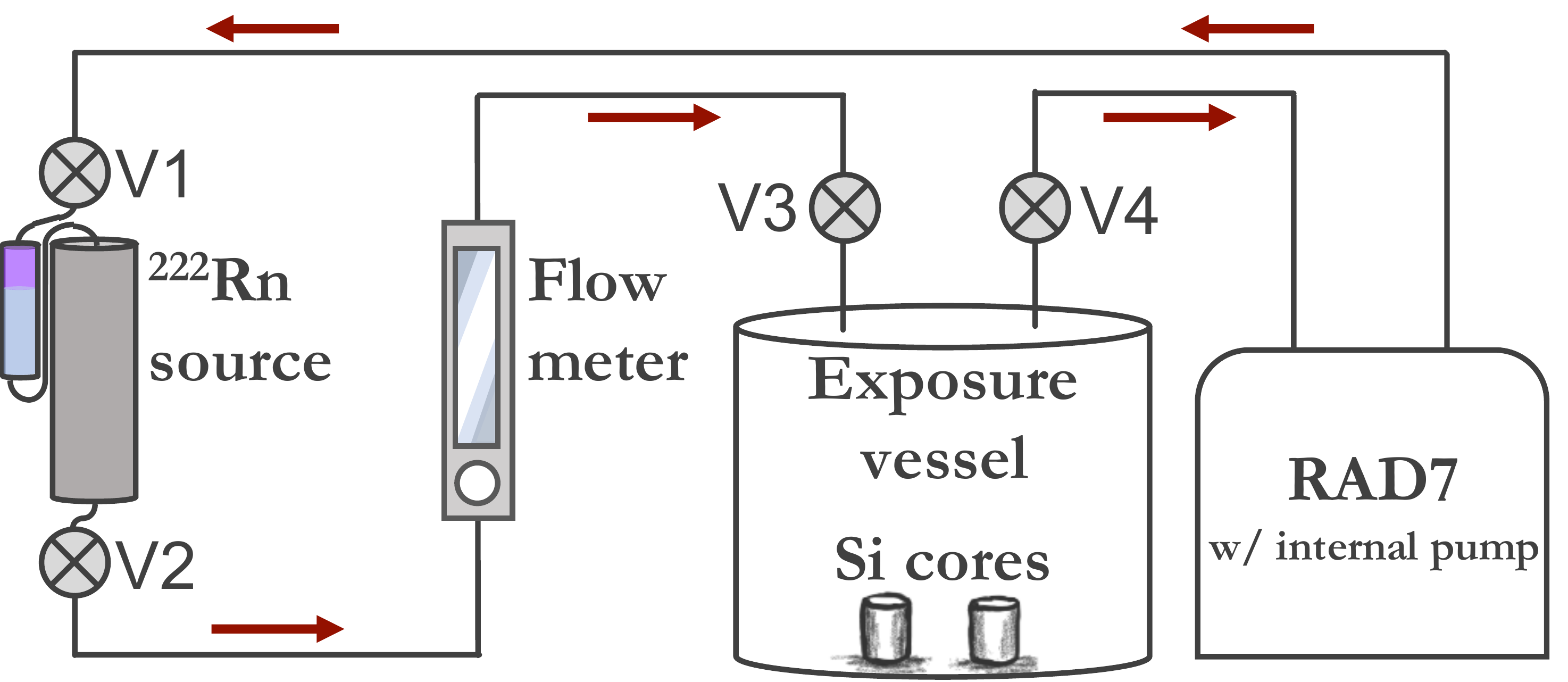}
  	\caption{During the second exposure, radon emanates from the $^{222}$Rn source into the air within the closed system. An inline flow meter indicates the circulation rate provided by the RAD7's internal pump. The RAD7 also measures the radon activity of the air passing through it. The valves V1 through V4 are open during the exposure, but may be closed afterward such that the high-radon air may be safely purged to an exhaust line (not shown). During the first exposure, the RAD7 in this figure is substituted with an inline hand-syphon pump.}
  	\label{f.expSetup}
\end{figure}

Two of the cores produced were used for this study. They were exposed to high-radon air (having a radon concentration of $\sim30\,$MBq/m$^3$) for 18 days and then another 13 days (with a 1-day assay of the $^{210}$Po decay rate in between). Figure~\ref{f.expSetup} illustrates the experimental setup during the last 13 days (the second exposure), which consisted of a $^{222}$Rn source supplied by Pylon~\cite{Pylon}, exposure vessel, inline flow meter, and Durridge RAD7 in a closed loop with the air inside circulated by the RAD7's internal pump. The 18-day exposure (the first exposure) used a hand-syphon pump in place of the RAD7.

\begin{figure}[!htbp]
	\centering
  	\includegraphics[width=1\linewidth]{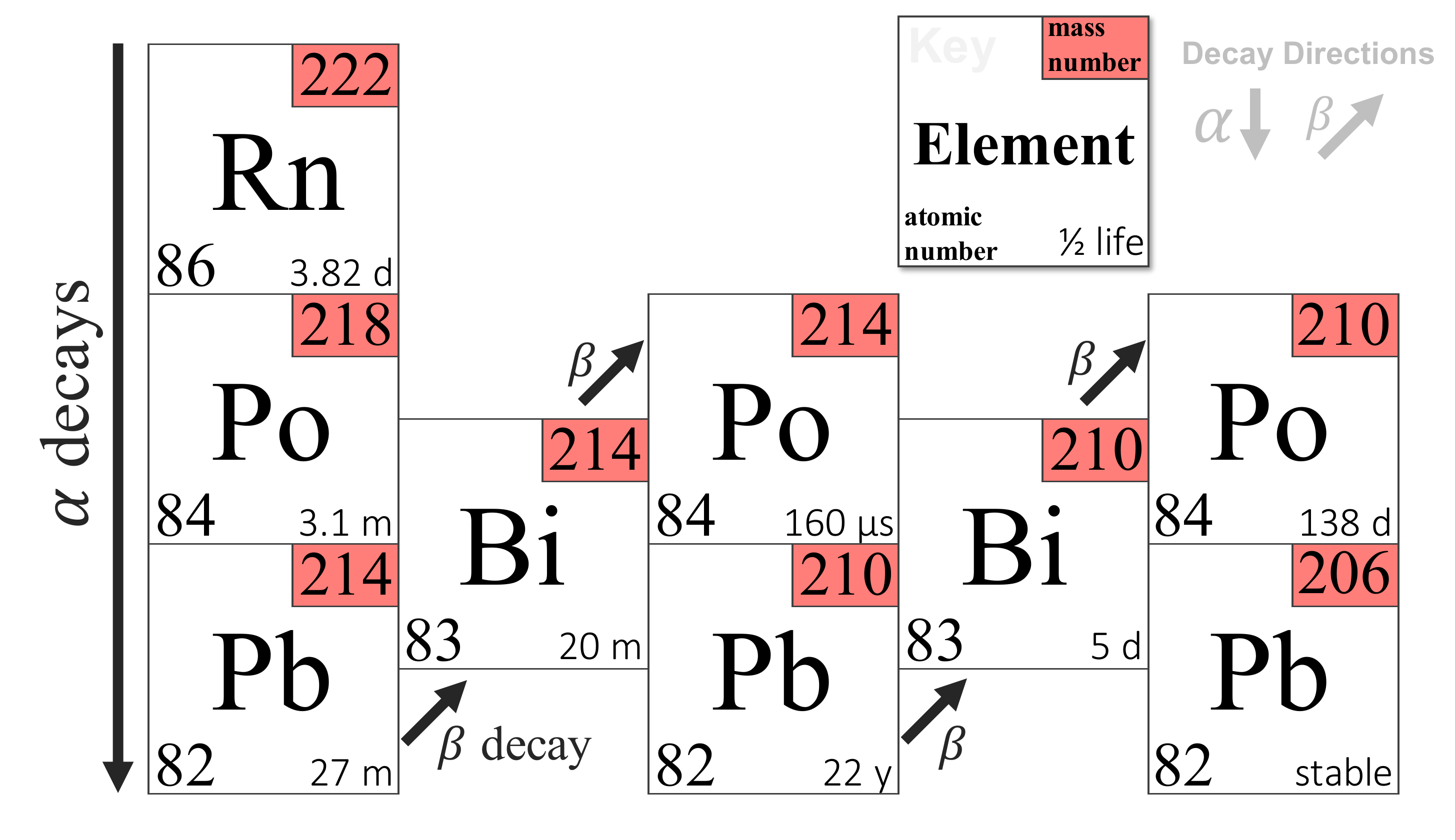}
  	\caption{$^{222}$Rn decay chain with key shows the atomic number, mass number, and half-life of elements. Radon and its progeny prior to $^{210}$Pb have relatively short half-lives. $^{210}$Pb has a long half-life of 22.3 years from which $^{210}$Po will grow into equilibrium over several months.}
  	\label{f.Rn222DecayChain}
\end{figure}

During exposure, $^{222}$Rn decays through its so-called fast daughters $^{218}$Po, $^{214}$Pb, $^{214}$Bi, and $^{214}$Po (see the radon chain in Fig.~\ref{f.Rn222DecayChain}). These fast daughters, usually positively charged~\cite{radonDaughters88percent}, may settle on nearby surfaces~\cite{Morrison_2018}. Subsequent alpha decays may then embed the daughters of order 10\,nm into the surface. As a result, some $^{210}$Pb is embedded and concentrated just below the surface from the sustained radon exposure.

Prior to each exposure, the setup was flushed with low-radon lab air. If sufficient mixing is obtained, the radon concentration within the exposure vessel is
\begin{align}
	C_\mathrm{Rn}(t) = \frac{A_\mathrm{source}}{V_\mathrm{source} + V_\mathrm{vess}}\left[ 1 - \exp{\left(-\frac{t}{\tau_\mathrm{Rn}}\right)}\right],
	\label{eq.C_Rn}
\end{align}
where the $^{222}$Rn source activity $A_\mathrm{source} = 118\pm5\,$kBq and its inner volume $V_\mathrm{source}= 0.2\,$liters. The volume of the exposure vessel $V_\mathrm{vess} = 3.5\,$liters and the radon mean lifetime $\tau_\mathrm{Rn} = 5.51\,$days. The exponential factor reflects radon growing into equilibrium within the setup volume.

During the first exposure, a syphon pump was used to circulate the gas. For the second exposure, the syphon pump was replaced by a RAD7, making use of its internal pump. The RAD7 demonstrated that sufficient mixing was achieved, so the time dependence of the radon concentration during the second exposure should be described by Eq.~\ref{eq.C_Rn}.

The relative amounts of $^{210}$Pb, $^{210}$Bi, and $^{210}$Po on the Si core at the beginning of the pre-etch assay influence the time dependence of the $^{210}$Po decay rate. If the radon concentration during exposure is understood, then the radioactive decay equations may be solved to provide the relative abundance of each radon daughter. Thus, the $^{210}$Po rate due to the second exposure can be modeled with a single fitting parameter accounting for the unknown absolute deposition rate. The $^{210}$Po rate due to the first exposure is characterized by three free parameters, representing the $^{210}$Pb, $^{210}$Bi, and $^{210}$Po at the end of the exposure, ensuring a conservative limit is set.

\begin{figure}[!htbp]
	\centering
  	\includegraphics[width=1\linewidth]{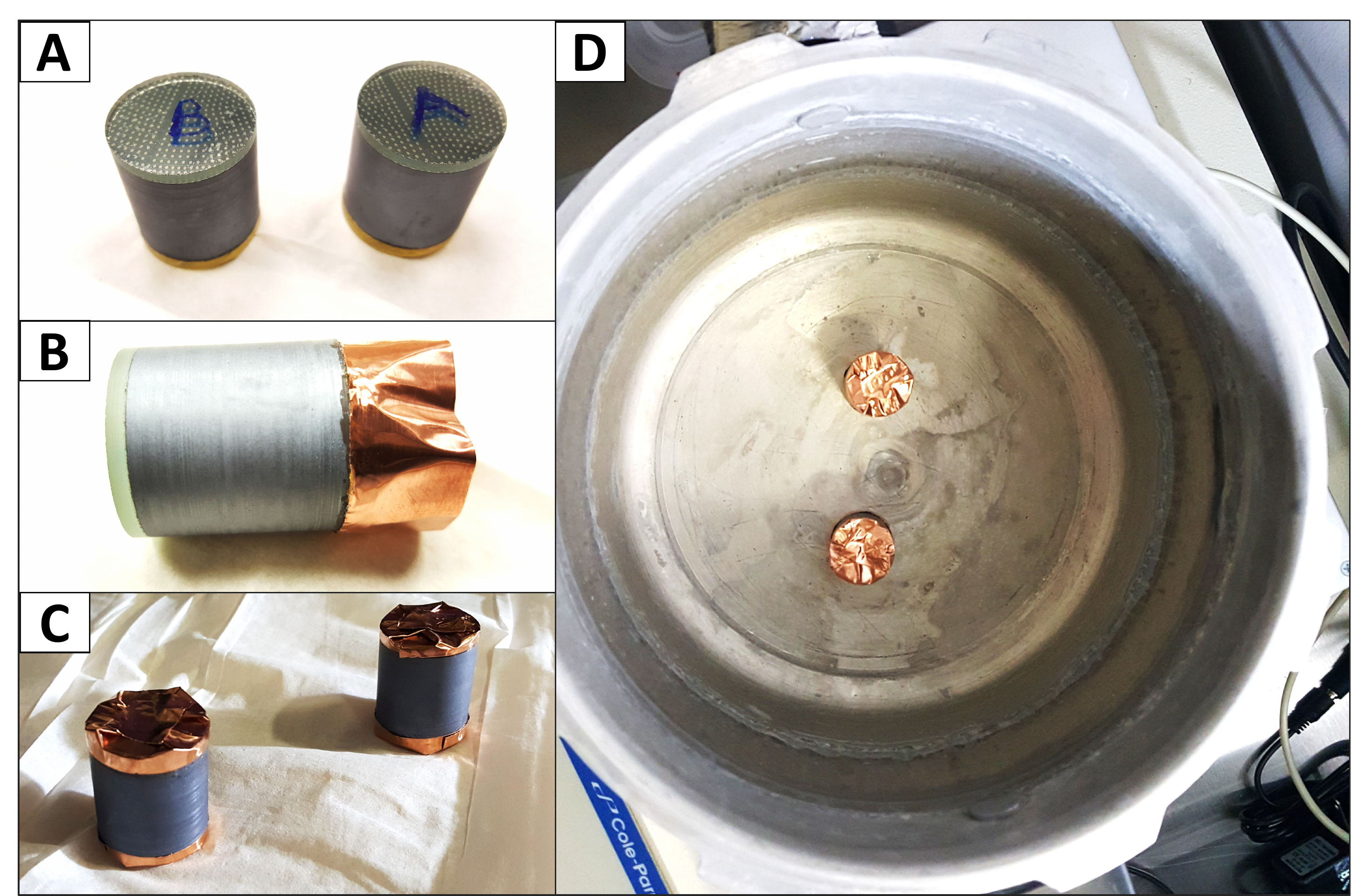}
  	\caption{Si core preparation and exposure geometry. \textbf{(A)}:~Wax on the faces protects delicate sensors from the etch and handling. \textbf{(B)}:~Copper foil shields charged daughters from the influence of static charge on the wax. \textbf{(C)}:~Si cores have copper foil in place and are ready for the high-radon exposure. \textbf{(D)}:~Si cores are placed roughly equidistant between the vessel walls. This configuration likely produces non-uniform radon-daughter deposition on core sidewalls.}
  	\label{f.expVessel}
\end{figure}

The exposure vessel was cylindrical with input and output feedthroughs penetrating its hemispherical lid. As seen in Fig.~\ref{f.expVessel}, copper foil was used to shield radon daughters from static charge on the wax that could reduce deposition on the core sidewalls. The cores were placed within the exposure vessel roughly equidistant between the vessel walls. The lack of radial symmetry in the core placement suggests non-uniform deposition with respect to the axial rotation of each core.

\section{Pre-etch Assay}
Though the pre-etch assay was performed on two Si cores that were both exposed to high-radon air, only the one that was eventually etched and re-assayed will be further discussed.

An Ortec Alpha Duo alpha spectrometer~\cite{Ortec:AlphaDuo} counted the $^{210}$Po rate on the sidewall of the Si core. The consistent placement and rotation of the Si core within the Alpha Duo detector bay was important to ensure that the same region of the core's sidewall was counted, as variation in the amount of $^{210}$Pb deposited as a function of rotation was likely, due to the placement geometry during exposure. Reproducibility of each angle of rotation minimized this possible systematic due to inadvertent rotation.

\begin{figure}[t]
	\centering
  	\includegraphics[width=1\linewidth]{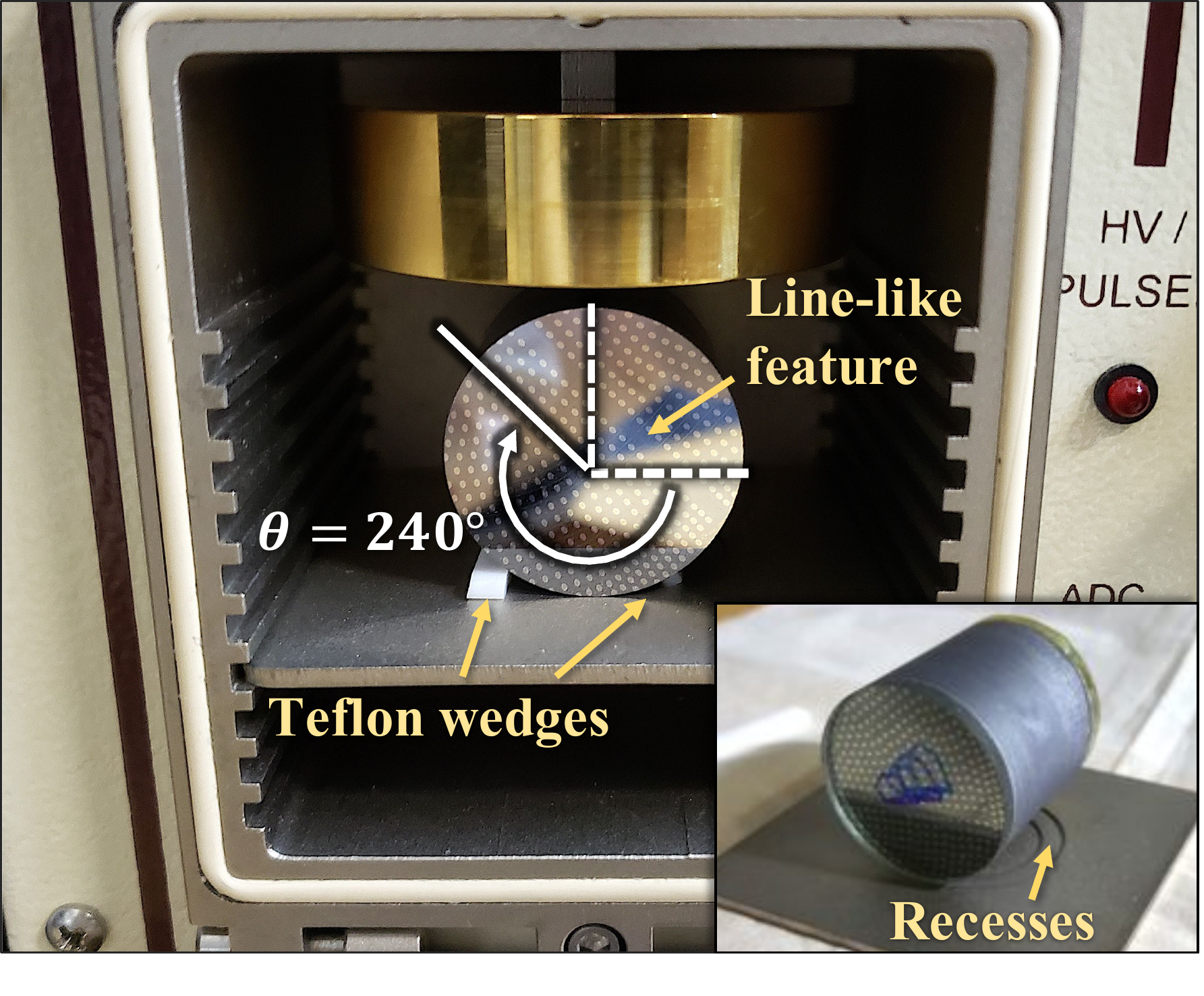}
  	\caption{Si core sitting inside the Alpha Duo counter detector bay. Low-rate teflon wedges ensure the core does not roll during measurement. A line-like feature between sensors on the core's face is used to track core rotation as measured clockwise from the horizontal. Inset shows the concentric recesses that were used to ensure consistent placement on the detector bay shelf.}
  	\label{f.SiCoreInBay}
\end{figure}

During the pre-etch assay, the Si core was periodically removed and replaced in the detector bay at an angle of rotation $\theta=0^{\circ}$, $120^\circ$, and $240^\circ$ as measured from the horizontal. The angle of the core was tracked by a line-like feature of the sensors on its face (see Fig.~\ref{f.SiCoreInBay}). Consistent core placement in the horizontal plane was guided by concentric recesses in the detector bay shelves. During handling, nitrile gloves were worn and only the detector faces were touched. This ensured that contaminants were neither added nor removed from the core's sidewall.

\subsection{Fitting the Pre-etch Data}
The Bateman equation $B_n$ describes the number of atoms of the $n$'th daughter produced by a radioactive parent~\cite{Bateman1910}:
\begin{align}
	B_n(t) = \sum_{i=1}^n{\left[ N_i\times \left(\prod_{j=1}^{n-1}{\lambda_j}\right) \times \sum_{j=i}^n{\left(\frac{\exp{\left(-\lambda_j t\right)}}{\prod_{p=i, p\neq j}^n{\left(\lambda_p - \lambda_j \right)}}\right)}\right]},
	\label{eq.BatemanEq}
\end{align}
where $\lambda_i$ is the $i$'th daughter's decay constant (for our case, $n=3$ and $i=1,2,3$ refers to $^{210}$Pb, $^{210}$Bi, and $^{210}$Po).

The two exposures are considered independently and, built upon Eq.~\ref{eq.BatemanEq}, added together. The expected signal representing the detected $^{210}$Po decay rate $S(t)$ has free parameters representing the initial $^{210}$Pb, $^{210}$Bi, and $^{210}$Po atoms at the end of the first exposure, abbreviated as $N_\mathrm{Pb1}$, $N_\mathrm{Bi1}$, $N_\mathrm{Po1}$ and the initial $^{210}$Pb at the end of the second exposure, abbreviated $N_\mathrm{Pb2}$:
\begin{align}
	S(t; N_\mathrm{Pb1}, N_\mathrm{Bi1}, N_\mathrm{Po1}, &N_\mathrm{Pb2}) \nonumber\\
	= \epsilon\lambda [B(t; N_\mathrm{Pb1}, &N_\mathrm{Bi1}, N_\mathrm{Po1})\nonumber\\
	 + \Theta & \left(t-\Delta t\right) B(t-\Delta t; N_\mathrm{Pb2})],
	\label{eq.fittingFunc}
\end{align}
where $B_{n=3}(t)$ is denoted as just $B(t)$ and $\Delta t$ is the time between the beginning of the first and second exposures. The detector efficiency is $\epsilon$, the $^{210}$Po decay constant $\lambda=5.813\times10^{-8}\,$s$^{-1}$, and the second term is multiplied by the Heaviside function.

\begin{figure}[!htbp]
	\centering
  	\includegraphics[width=\linewidth]{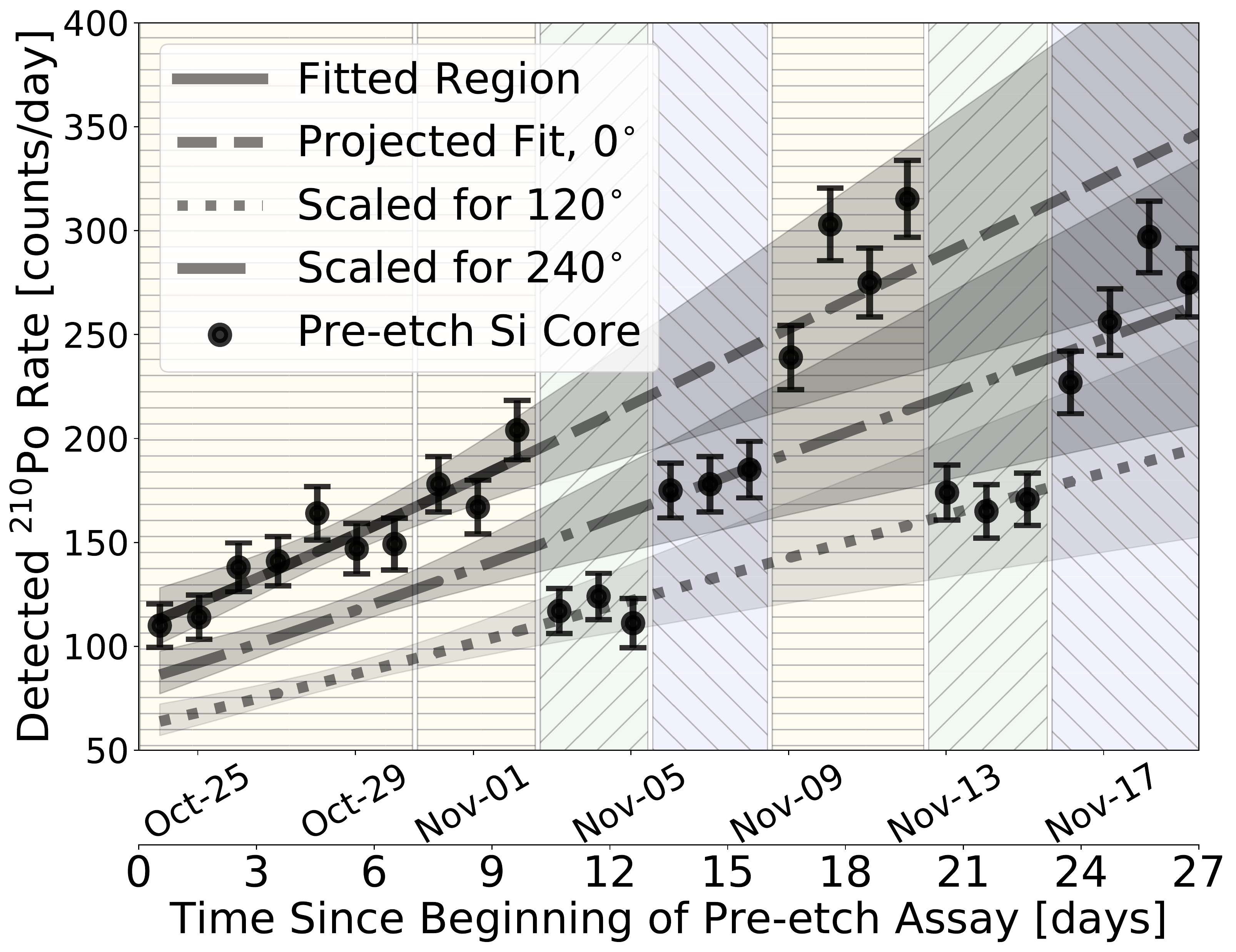}
  	\caption{Pre-etch data (dots with $1\sigma$ error bars) at varying core rotations. The hatch marks in the background of the plot indicate the $0^\circ$ (horizontal), $120^\circ$ (up to the right), and $240^\circ$ (down to the right) rotation of the core. The signal function was fit to the first $\theta=0^{\circ}$ data run (represented by the solid curve) with $\chi^2$/ndf $=7.3/9$, p-value $=0.7$. Its projection (dashed curve) agrees with the second $\theta=0^{\circ}$ run within statistical uncertainties on the best fit (upper shaded region). Days 11-13 show a lower rate for $\theta=120^{\circ}$ (dotted curve), as do days 14-16 for $\theta=240^{\circ}$ (dot-dashed curve), and each is consistent with its second run (see Table~\ref{t.angleDependence}). Both the $\theta=120^{\circ}$ and $\theta=240^{\circ}$ curves and their shaded uncertainty regions are found by scaling the best-fit $\theta=0^{\circ}$ curve. Not shown is the earlier one-day assay, which measured $21\pm6$ counts per day (cpd) compared to the best-fit $21^{+12}_{-8}$ cpd for that time.}
  	\label{f.exposure}
\end{figure}

Equation~\ref{eq.fittingFunc} was fit to the ($\theta = 0$) one-day assay after the first exposure and the first 10 days of the ($\theta = 0$) assay after the second exposure. Figure~\ref{f.exposure} shows this fit, but focused on the pre-etch assay after the second exposure for clarity. The inferred surface contamination $\Sigma$ on the measured ($\theta=0$) side of the Si core at the beginning of the pre-etch assay (prior to etch) was
\begin{align}
	\Sigma_\mathrm{Pb} &= 137^{+53}_{-60}\,\mathrm{Bq/m}^2,\nonumber\\
	\Sigma_\mathrm{Bi} &= 101^{+25}_{-23}\,\mathrm{Bq/m}^2,\nonumber\\
	\Sigma_\mathrm{Po} &= 7^{+1}_{-1}\,\mathrm{Bq/m}^2.\nonumber
\end{align}

Uncertainties were determined by throwing $10^{7}$ unique fitting parameter combinations each drawn from a uniform distribution. A $1\sigma$ cut on the resulting $\chi^2$ was placed to obtain $1\sigma$ bands. The domain of each thrown parameter and the best-fit value is shown in Table~\ref{t.ThrownDomain}.
\begin{table}[H]
\centering
\begin{tabular}{r | *{2}{c}}
	parameter & domain & best-fit value \\
	\hline
	$N_\mathrm{Pb1}$ & $[0,\>2\times10^{7}]$ & $1.1\times10^{7}$ \\
	$N_\mathrm{Bi1}$ & $[0,\>2\times10^{4}]$ & $4.2\times10^{3}$ \\
	$N_\mathrm{Po1}$ & $[0,\>2\times10^{4}]$ & $4.2\times10^{3}$ \\
	$N_\mathrm{Pb2}$ & $[0,\>2\times10^{8}]$ & $1.4\times10^{7}$ \\
\end{tabular}
\caption{Domains for each thrown parameter in units of the number of atoms. The best-fit values were found by simultaneous $\chi^2$ minimization.}
\label{t.ThrownDomain}
\end{table}
Of the $10^{7}$ parameter combinations thrown, 21,108 passed the $1\sigma$ cut. These values were checked to ensure good coverage of the parameter space.

Each run at each angle of core rotation was fit, with a single scaling parameter, against the signal function fit to the first $0^\circ$ run. That is, the $0^\circ$ fit of $S(t)$, was scaled to fit each run as
\begin{align}
	F(t) = A\times S(t),
	\label{eq.scaledFit}
\end{align}
where $A$ is a best-fit scaling parameter. Table~\ref{t.angleDependence} lists values of $A$ and its uncertainty $\sigma_A$ for both runs at each core rotation. The difference between the two scaling factors for each angle is consistent with statistical uncertainties.
\begin{table}[H]
\centering
\begin{tabular}{r | *{3}{r}}
	$\theta$ & $0^\circ$ & $120^\circ$ & $240^\circ$ \\
	\hline
	$A_1$ & 1.00 & 0.56 & 0.76 \\
	$A_2$ & 1.05 & 0.57 & 0.77 \\
	$\sigma_{A_1}$ & 0.02 & 0.03 & 0.06 \\
	$\sigma_{A_2}$ & 0.05 & 0.02 & 0.03 \\
	$|A_1 - A_2|$ & 0.05 & 0.01 & 0.01 \\
\end{tabular}
\caption{Angle dependence and reproducibility. $A_1$ and $A_2$ represent the scaling factors for the first and second runs at each angle relative to the first $0^{\circ}$ run, with $\sigma_{A_1}$ and $\sigma_{A_2}$ their respective fit uncertainties. The difference between these scaling factors $|A_1 - A_2|$ for each angle is consistent with statistical uncertainties.}
\label{t.angleDependence}
\end{table}

\section{Etching Technique}
After high-radon exposure, the Si core received a heavy etch treatment with a mixture of 80\% nitric acid, 16\% hydrofluoric acid, and 4\% acetic acid within a large Nalgene tub. The substrate diameter was measured to be 25.446$\pm$0.001\,mm prior to etching. The substrate was then placed into a PTFE basket and dunked into the acid solution for 30 seconds. Following the acid dunk, the substrate was dunked and rinsed with DI water. The substrate diameter was again measured and found to be 25.430$\pm$0.001\,mm, indicating that 16$\pm$1\,\si{\um} was removed---much more than the $\sim$0.1\si{\um} implantation depth of radon daughters.

The glass was removed from both sides of the Si core by heating it to 100\,$^\circ$C and sliding the glass away. Placing the Si core in a beaker with high-purity trichloroethylene removed the remaining wax. The Si core was then rinsed with DI water and cleaned using isopropanol, acetone, methanol, and then blown dry with nitrogen.

\section{Post-etch Assay and Observed $^{210}$Pb Reduction}
\begin{figure}[t]
  	\centerline{\includegraphics[width=1\linewidth]{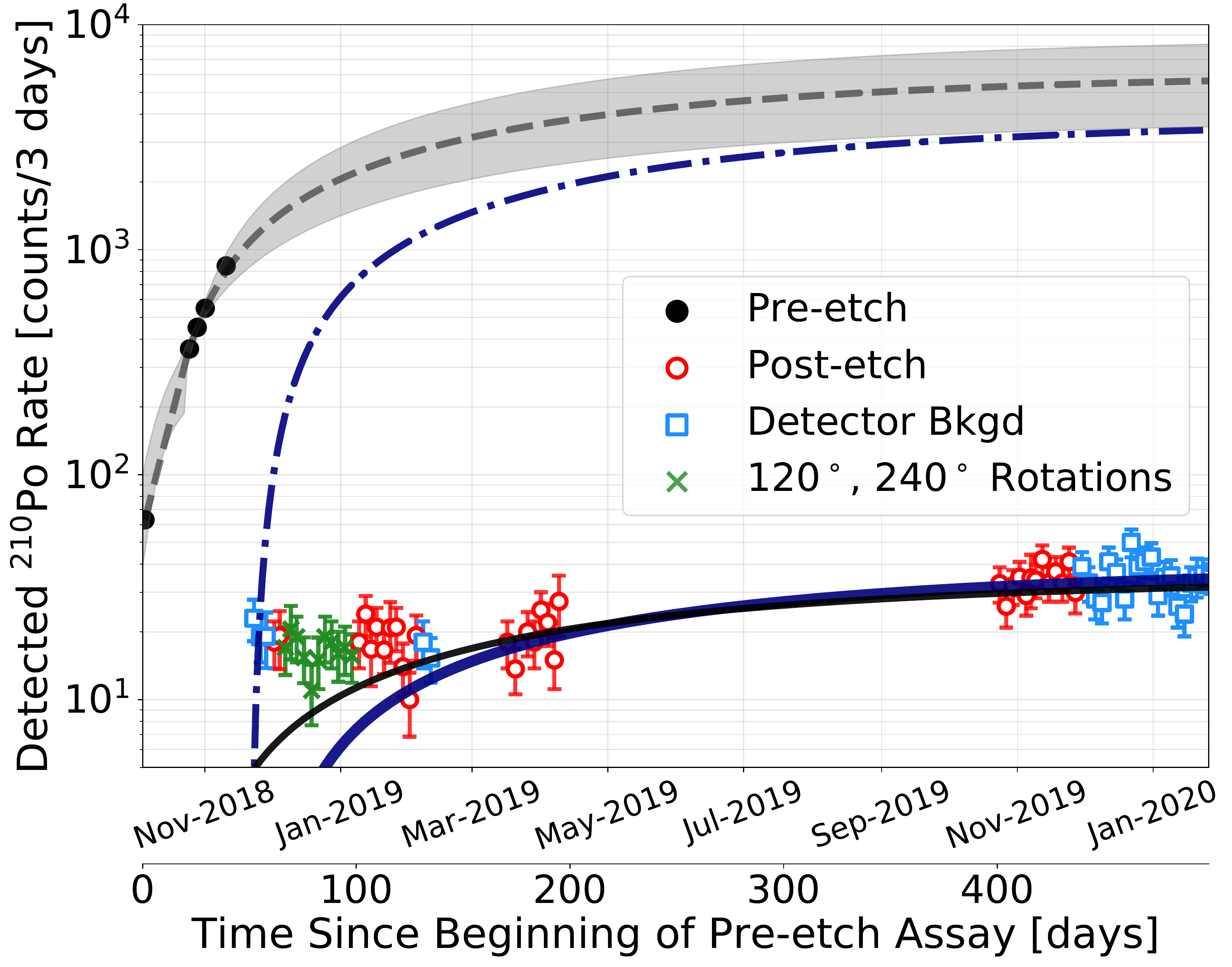}}
  	\caption{The signal function (Eq.~\ref{eq.fittingFunc}) is fit (dashed line) to the pre-etch $^{210}$Po rate (solid black circles).  Its projection shows the expected grow-in from implanted $^{210}$Pb with its 1$\sigma$ band shown in light gray around it. Applying the Optimum Interval method~\cite{Yellin:aa,Yellin:2007aa} to the post-etch $^{210}$Po rate (red circles) using the 1$\sigma$ lower limit of the signal function projection results in a scaled function at 90\% confidence level upper limit (solid black curve) that corresponds to a reduction of (at least) 110$\times$. In the case that the etch removes all $^{210}$Bi and $^{210}$Po, the 1$\sigma$ lower limit on the signal after etch changes (navy dot-dashed curve) and produces a 90\% upper limit (thick navy curve) corresponding to a reduction of (at least) 99$\times$. The post-etch assay agrees with background measurements (blue squares) taken early and late in the post-etch assay (indicating an increase in the detector background with time). Though not used for this reduction measurement, $\theta=120^\circ$ and $\theta=240^\circ$ runs are also shown (green $\times$'s).}
  	\label{f.pre_post_assay}
\end{figure}

While the Si core sidewall was being etched, a background measurement was made with the detector unchanged except that the core was not present. After the etch was finished, the core sidewalls were again counted. Figure~\ref{f.pre_post_assay} shows the pre- and post-etch assays. Again, between post-etch runs, a background measurement was made and showed that the post-etch rate was consistent with the background rate. Though the $\theta=0^{\circ}$ angle was the primary focus, $\theta=120^\circ$ and $\theta=240^\circ$ rates were also measured and were also consistent with the background rate.

The Optimum Interval (OI) method allows a limit to be set, with or without knowledge of the background, providing the expected signal is understood~\cite{Yellin:aa,Yellin:2007aa}. After etching, the expected signal is still $S(t)$, but scaled by a reduction factor that represents the effect of the sidewall etch. The OI method returns the reduction factor $R \equiv 1/A$, as defined by Eq.~\ref{eq.scaledFit}, that can be stated at 90\% confidence limit (C.L.). By this method, a reduction of $R>110$ at 90\% C.L. on the 1$\sigma$-lower-limit $^{210}$Po rate was found, if all isotopes are removed equally.

The removal of $^{210}$Pb, $^{210}$Bi, and $^{210}$Po from the surface of Ge was similarly studied by M. W\'ojcik and G. Zuzel~\cite{W_jcik_2012}. They found that $^{210}$Po ($^{210}$Pb) was removed more efficiently from rougher (smoother) surfaces, by almost a factor of 10. Because the Si core sidewall assayed in this study would be considered a rough surface, $^{210}$Po was likely preferentially removed.

If $^{210}$Po was indeed preferentially removed by this etch, the expected signal would then represent $^{210}$Po growing in from the $^{210}$Pb not removed by the etch. By assuming all $^{210}$Bi and $^{210}$Po are removed, this analysis indicates, rather conservatively, a $^{210}$Pb reduction of $R>99$ at 90\% C.L.

Though the OI method can account for knowledge of the background and would likely provide a stronger limit on reduction, the background rate has not been used. This is because it is not yet known how the background rate changes with the Si core inside the Alpha Duo detector bay. Some line-of-sight backgrounds are blocked and this would need to be better understood. In addition, the observed reduction is already sufficiently large that, for this study, additional work is unwarranted.

\section{Conclusions}
By comparing the $^{210}$Po rates before and after the sidewall-etching technique was performed, a reduction of $R>99$ at 90\% C.L. was found for $^{210}$Pb. This limit assumes all $^{210}$Bi and $^{210}$Po were removed by the etch thus making the statement of $^{210}$Pb reduction conservative.

The application of this sidewall-etching technique applied to SuperCDMS SNOLAB, for example, would reduce the detector sidewall surface contamination, directly improving the experiment's sensitivity. The technique also provides a method to salvage a detector unintentionally subjected to high rates of radon-daughter deposition.

\section*{Acknowledgements}
We would like to thank Dr. Robert Calkins for his suggestions on this work.

Funding: This work was supported by the Department of Energy [grant numbers DE-SC0014036, DE-SC0017859, DE-AC02-76SF00515, and DE-SC0014223]; the National Science Foundation [grant number PHY-1506033].

\section*{References}

\bibliography{SidewallEtchPaper_arxiv_v16.bib}

\end{document}